\documentclass{appolb}
\usepackage{graphicx}
\usepackage{lineno, blindtext}
\usepackage{cite}
\usepackage{subcaption}

\begin{document}

\title{Heavy Flavour measurements in Pb--Pb collisions with the upgraded ALICE Inner Tracking System 
\thanks{Presented at Excited QCD 2020}%
}
\author{D. Andreou on behalf of the ALICE Collaboration
\address{CERN, 1211 Geneva 23, Switzerland, \\Nikhef, Amsterdam, The Netherlands}
\\
}
\maketitle
\begin{abstract}
ABSTRACT: During the second LHC long shutdown (LS2) the Inner Tracking System (ITS) of ALICE (A Large Ion Collider Experiment) will be replaced by seven layers of CMOS Monolithic Active Pixel Sensors (MAPS). The latest innovations in silicon imaging technology allow for the construction of large, ultra-thin silicon wafers which can further improve the capabilities of the ALICE tracker. The research and development studies towards the construction of a novel vertex detector have started. The detector installation has been proposed for the third LHC long shutdown (LS3) during which the three innermost layers shall be replaced by three cylindrical layers of large curved CMOS wafers. This upgrade (ITS3) will further improve the impact parameter resolution and the tracking efficiency of low momentum particles. The innermost layer will be positioned closer to the interaction point and the material budget will be reduced down to 0.05~$\%X_0$ per layer.\\
Monte Carlo simulations of a simplified ITS3 geometry within the ITS2 design indicate an improvement in the impact parameter resolution and the tracking efficiency, which are of crucial importance for measurements of heavy-flavour hadrons. This contribution shows the improved performance for the example of the $\Lambda_\textnormal{b}$, for which the significance of its measurement is extracted based on these MC simulations. A significant improvement by almost a factor of three in the low momentum region compared to the ITS2 is observed.

\end{abstract}

\section{Introduction}
The ALICE experiment at the CERN Large Hadron Collider (LHC) is studying the properties of the Quark Gluon Plasma (QGP). Important information can be extracted by studying the behaviour of heavy flavour particles after their journey in the QGP. Future opportunities on high-density QCD measurements after LHC long shutdown 2 (LS2) have been presented \cite{a}. The main topics that will be explored through the heavy flavour measurements are the characterization of the macroscopic long wavelength QGP properties with high precision and the investigation the parton dynamics responsible for the QGP properties. More specifically, the focus will be on putting strong constraints on the transport coefficients of heavy quarks which will help to identify the leading interaction mechanisms of heavy quarks and derive their level of thermalization. This will be done through high precision measurements of transverse momentum anisotropies and of the nuclear modification factors of heavy flavoured hadrons, which are used to constrain the heavy quark diffusion coefficient and its dependence on temperature. The new measurements of D and B mesons and of charm and beauty baryons will not only give important information about the thermalization of heavy quarks in the medium but will also indicate to which degree the process of recombination of heavy quarks with lighter quarks contributes to the hadronisation process. Recent studies have shown that recombination is expected to be dominant in the low momentum region in Pb--Pb collisions since the increased amount of partons allows the heavy quarks to hadronise with light ones through recombination \cite{a}. The hadronisation through recombination results in the increased production rate of many heavy flavour species, the measurements of which are currently limited by statistics. With the increase of the instantaneous luminosity in the following years and with the usage of cutting edge detector technologies many heavy flavour measurements will be improved and new ones will be made possible.

\section{The ALICE Inner Tracking System upgrades}
\begin{figure}[!htbp]
\centerline{
\includegraphics[width=4.7cm]{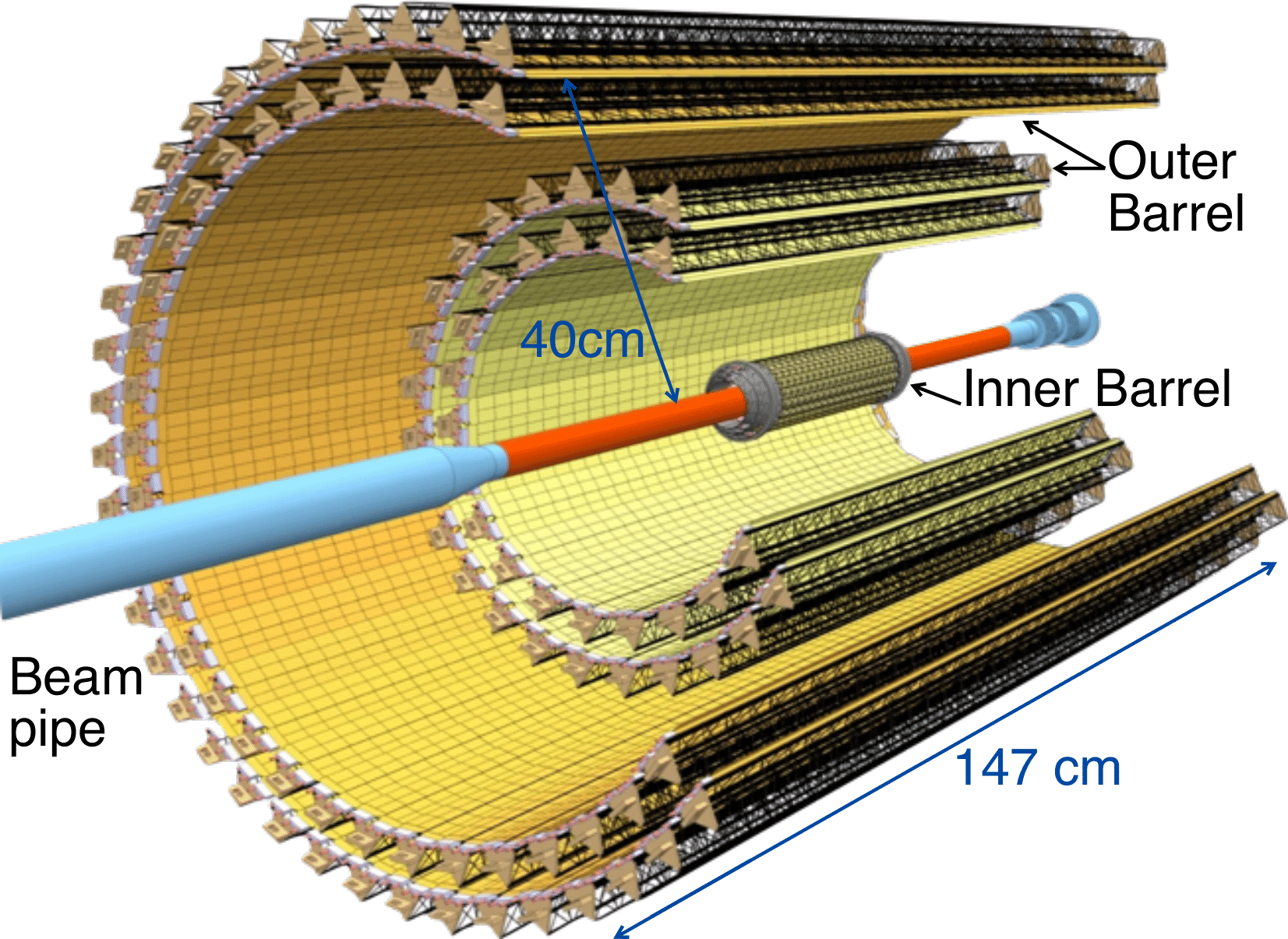}}
\caption{Detector overview of the ALICE ITS2. Copyright CERN, reused with permission.}
\label{Fig:ITS2}
\end{figure}

During LS2 a major upgrade of the ALICE Inner Tracking System is taking place. The 6 layers of the ITS1 with three different technologies (pixel, drift, strip) will be replaced by 7 layers of CMOS Monolithic Active Pixel Sensors (MAPS) as demonstrated in the detector overview design of ITS2 in Figure \ref{Fig:ITS2} \cite{b}. The small thickness of the MAPS (50~$\mu m$ in the inner barrel) along with other modifications in the mechanical design reduce the material budget from 1.14$\%$~$X_0$ down to 0.35$\%$~$X_0$ per layer in the Inner Barrel. The innermost layer of the ITS is placed closer to the interaction point and the continuous read-out rate capability will be increased from 1~kHz to 100~kHz in Pb--Pb collisions. All the above modifications contribute to the improvement of the impact parameter resolution and the tracking efficiency in the low momentum region.

In order to further improve the tracking capabilities during Run 4, ALICE has started a research and development effort towards a further upgrade of the ITS, the ITS3, during the third LHC long shutdown \cite{c}. During the ITS3 upgrade the three innermost layers of the ITS2 will be replaced by three ultra-light cylindrical layers of silicon pixel detectors. In an effort to reduce further the material budget, large scale ultra-thin silicon wafers will be produced through the process of stitching and will operate as single sensors. The silicon wafers will be thinned down to 20-40~$\mu$m, a thickness at which the flexible nature of silicon allows the bending of the silicon wafers. The wafers will be wrapped around the beam pipe. The periphery of the sensor, the interface responsible for the sensor configuration and serial data, will be positioned outside the fiducial volume eliminating the need of external devices in the active volume.

\begin{figure}[!htbp]
\centerline{
\includegraphics[width=6.9cm]{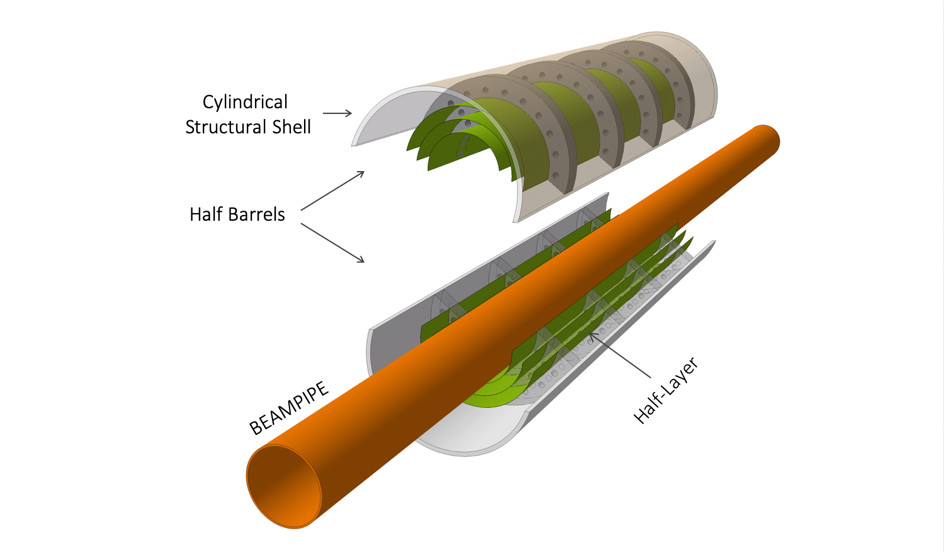}}
\caption{Detector design of the inner layers of the third ALICE Inner Tracking System. Copyright CERN, reused with permission.}
\label{Fig:ITS3}
\end{figure}

\begin{figure}[!htbp]
\begin{subfigure}{.5\textwidth}
\centering
\includegraphics[width=6.cm]{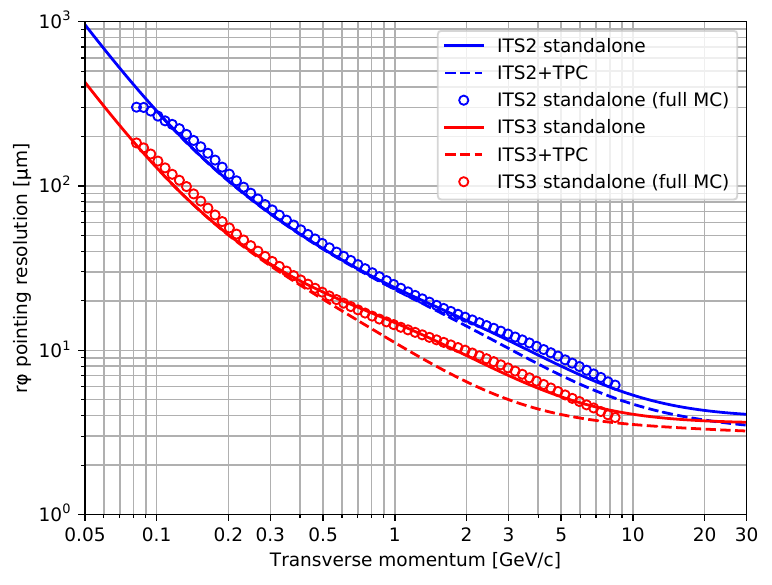}
\end{subfigure}
\begin{subfigure}{.5\textwidth}
\centering
\includegraphics[width=6.cm]{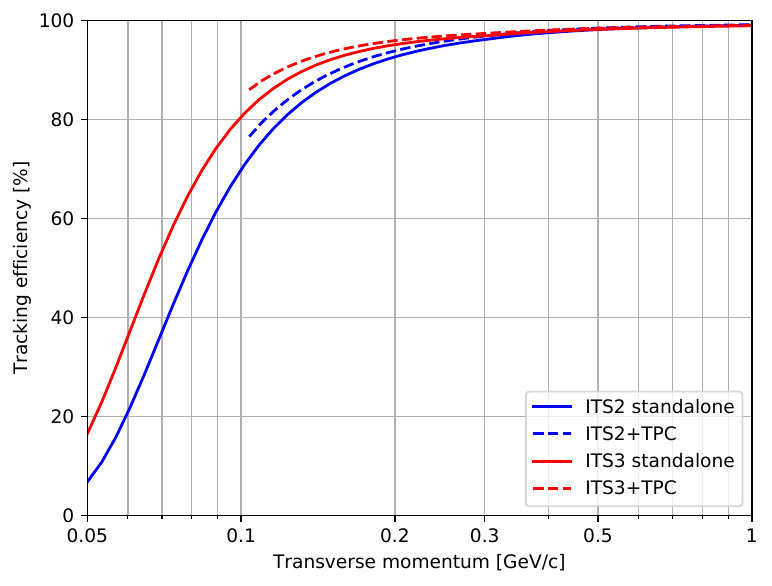}
\end{subfigure}\caption{Pointing resolution in r$\phi$ plane (left) and tracking efficiency (right) with the ITS2 (blue) and the ITS3 (red). Copyright CERN, reused with permission.} 
\label{Fig:res}
\end{figure}

In this way in the area of interest there will be only silicon and low mass open-cell carbon foam cells for the mechanical support of the sensors (Fig. \ref{Fig:ITS3}). The water cooling that is present in the ITS2 will be replaced by air cooling, possible by reducing the power consumption of the sensors to 20~mW/c$m^2$. With all the above modifications the material budget will decrease from 0.35$\%$~$X_0$ to 0.05$\%$~$X_0$ per layer. Additional modifications will be done on the beam pipe, the thickness of which will be reduced from 800~$\mu$m to 500~$\mu$m and the inner radius from 22~mm to 18~mm. This will allow the innermost layer of the ITS to be moved closer to the interaction point. The detector upgrade along with the installation of the new beam pipe will improve the impact parameter resolution and the tracking efficiency in the low momentum regions, as indicated in Figure \ref{Fig:res}. The pointing resolution and the tracking efficiency were studied with a Fast Monte Carlo Tool (FMCT) which takes into account multiple scattering, secondary interactions and detector occupancy, but ignores the energy loss of the particle in the detector and in the beam pipe. The pointing resolution and the tracking efficiency were estimated considering a standalone ITS (solid lines) and the combined case of the ITS plus the Time Projection Chamber (TPC) (dashed lines) for ITS2 and ITS3. The distribution of the pointing resolution indicates an improvement by a factor two for $p_\mathrm{T}\simeq1GeV/c$ with the ITS3+TPC information. The results of the FMCT were confirmed by a full Monte Carlo study that was done for a standalone ITS2 and ITS3 and is demonstrated in the plot with circles. 

\section{Prospects on heavy flavour measurements}

The better pointing resolution and tracking efficiency offered by the ITS3 will improve many heavy flavour measurements that are currently limited and enable new ones. Of particular interest is the study of the enhancement of the baryon to meson ratios of many heavy flavour species expected from recombination observed in Pb--Pb collisions compared to p-p collisions. Preliminary studies have shown an enhancement of the $\Lambda_c / D^0$ and $D_s / D^0$ ratios and there are predictions for other strange mesons and beauty baryons, namely for the $B_s$ and $\Lambda _b$ \cite{a}. However, the measurements that quantify the degree of the enhancement are limited for most of the species. Based on MC simulations it was demonstrated that with the ITS3 the measurements of all the above particles will be improved compared to ITS2. 

$\Lambda _b$ is a beauty baryon with udb quarks and mass 5.62 $GeV/c^2$ \cite{d}.  The decay channel that was studied is the $\Lambda^{0} _b \rightarrow \Lambda^{+} _c + \pi^{-}$ with $ \Lambda^{+} _c  \rightarrow  \pi^{+} + p + K^{-}$. The branching ratio (BR) of the $\Lambda _b$ in this channel of interest is very small, $BR_{\Lambda^{0} _b}=0.49\%$, $BR_{\Lambda^{+} _c }=6.28\%$,  and the combinatorial background very large since it is a four prong final state. In order to be able to reconstruct and separate the signal, it is very important to have improved vertex resolution and to follow a tight topological selection \cite{b}. The same analysis strategy was followed for the ITS2 and the ITS3 cases for the calculation of significance $Signal \over Signal + Background$  of measuring the $\Lambda _b$  particle. 

\begin{figure}[!htbp]
\begin{subfigure}{.5\textwidth}
\centering
\includegraphics[width=6.5cm]{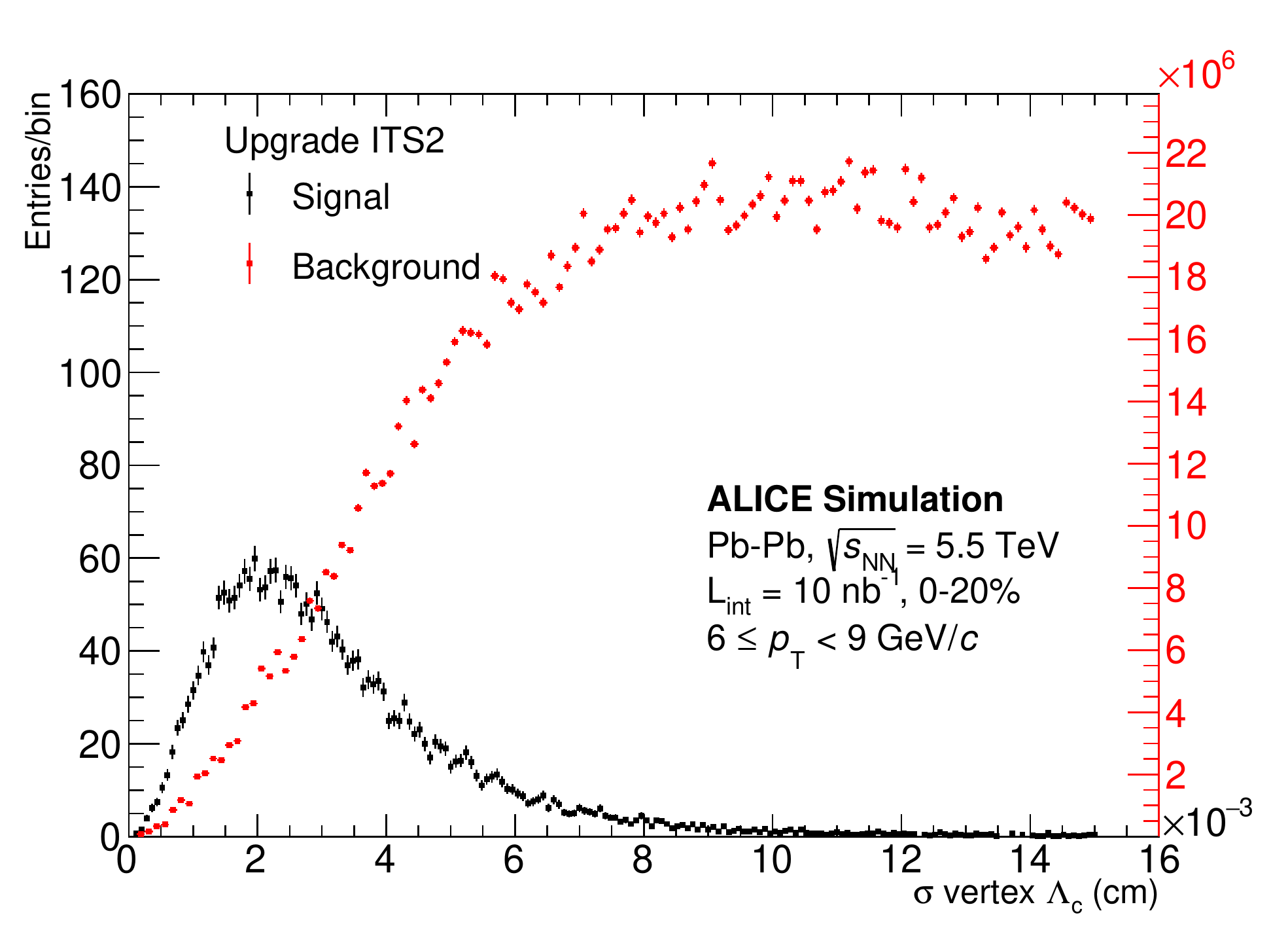}
\end{subfigure}
\begin{subfigure}{.5\textwidth}
\centering
\includegraphics[width=6.5cm]{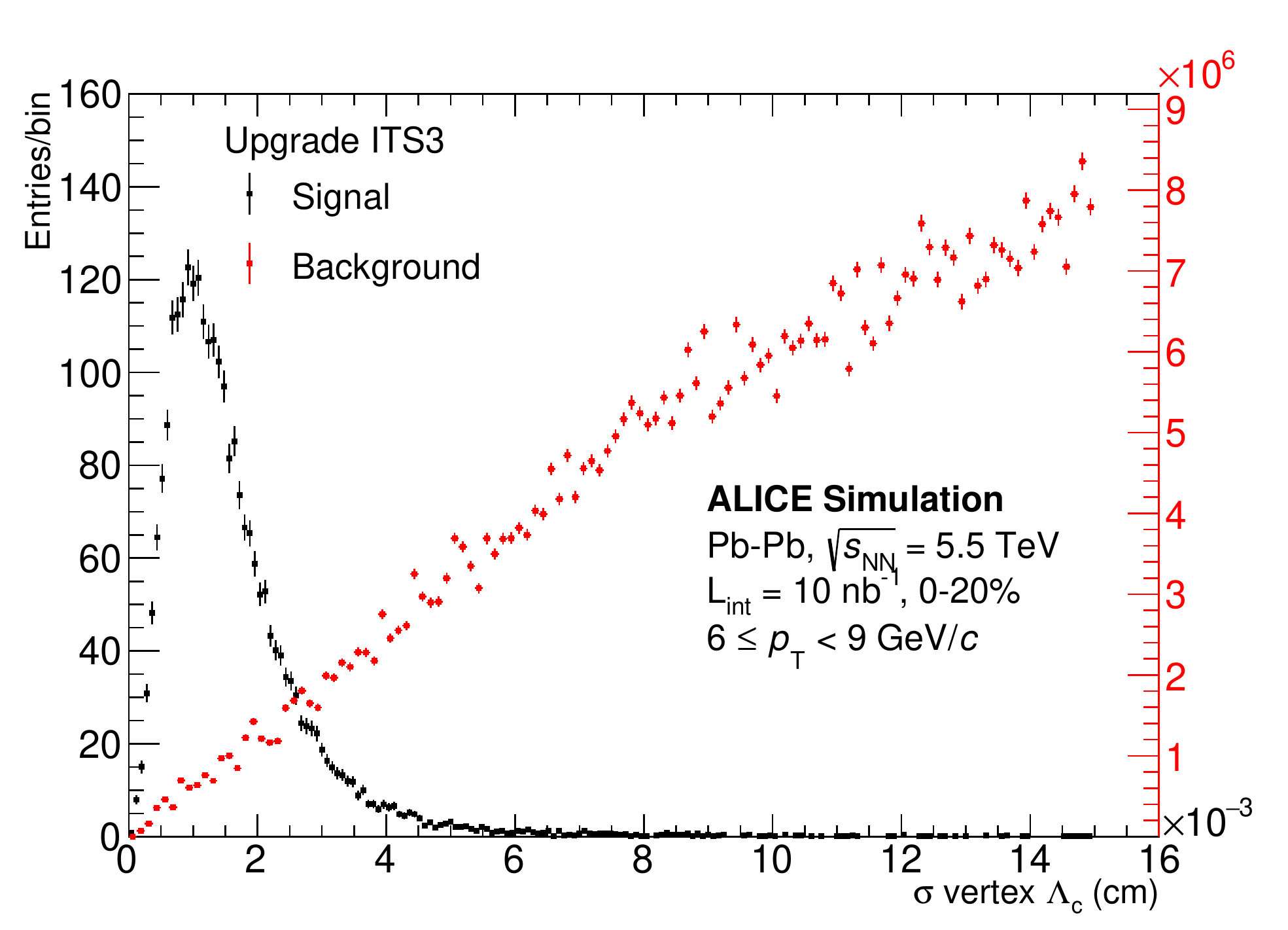}
\end{subfigure}
\caption{Signal (black) and background (red) distributions of the $\sigma$ vertex $\Lambda _c$ with the ITS2 (left) and the ITS3 (right). Copyright CERN, reused with permission.}
\label{Fig:sigmavtx}
\end{figure}

\begin{figure}[!htbp]
\centerline{
\includegraphics[width=6.5cm]{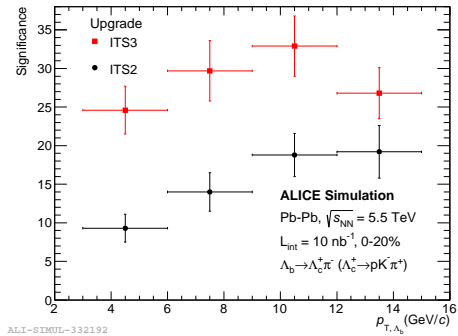}}
\caption{Significance measurement of $\Lambda _b$ particle with the ITS2 (black) and the ITS3 (red). Copyright CERN, reused with permission.}
\label{Fig:significance}
\end{figure}

The distributions of the topological variables of the decay were studied for both cases and optimal cuts were selected in order to keep the signal and reject a significant amount of background for the maximization of the significance. As shown in Fig.~\ref{Fig:sigmavtx} the effect of the improvement is very large in the signal and background distributions of the track dispersion around the decay vertex, the $\sigma$ vertex of $\Lambda _c$. The signal distribution is narrower making the signal to background separation in the ITS3 case larger and allowing for tighter cuts in the selection. The significance of measuring the $\Lambda _b$  in 0-20$\%$ centrality at 5.5 TeV in Pb--Pb collisions was calculated in four $p_T$ bins and scaled to luminosity $L=10~\mathrm{nb^{-1}}$(further assumptions are detailed in the reference) \cite{e}. As shown in Fig.~\ref{Fig:significance} with the ITS3 the significance of measuring the $\Lambda _b$ particle will significantly increase, especially in the low momentum region where it is estimated to improve by almost a factor of three.

\section{Conclusions}
The upgrade of the ALICE ITS3,  with the latest innovations in silicon imaging technology, will improve the impact parameter resolution and the tracking efficiency of low momentum particles. The effect of these improvements is reflected by estimates for the measurements of the heavy flavour particles, strange mesons and charm and beauty baryons, in a wide momentum range. The significantly improved reconstruction will enable high precision measurements of transverse momentum anisotropies and of nuclear modification factors, revealing new information on the processes of hadronization and thermalization of heavy quarks.

\end{document}